\documentclass{aastex}
\usepackage{emulateapj5}
\usepackage{psfig}

\begin{document}
\title{Some Constraints On the Effects of Age and Metallicity on the Low Mass X-ray Binary Formation Rate\altaffilmark{1}}

\author{Arunav Kundu\altaffilmark{2}, Thomas J. Maccarone\altaffilmark{3}, Stephen E. Zepf\altaffilmark{2}, \& Thomas H. Puzia\altaffilmark{4} }

\altaffiltext{1} {Based on observations made with the NASA/ESA Hubble Space Telescope, obtained at the Space Telescope Science Institute, which is operated by
    the Association of Universities for Research in Astronomy, Inc., under NASA contract
    NAS 5-26555, and on observations made with the Chandra X-ray Observatory.}
\altaffiltext{2}{ Physics \& Astronomy Department, Michigan State University, East Lansing, MI 48824. e-mail: akundu, zepf @pa.msu.edu}

\altaffiltext{3}{ SISSA/ISAS, via Beirut n. 2-4, 34014, Trieste, Italy. e-mail: maccarone@ap.sissa.it}

\altaffiltext{4}{Sternwarte der Ludwig-Maximilians-Universit\"at, Scheinerstrasse 1, 81679 M\"unchen, Germany puzia@usm.uni-muenchen.de}

\begin{abstract}
We have studied the low mass X-ray binary (LMXB) populations within and outside globular clusters (GC) in NGC 4365 and NGC 3115. Using published age and metallicity constraints from optical and IR observations of their GCs, we do not find any evidence for an increase in the LMXB formation rate in the intermediate age cluster population of NGC 4365, as has been proposed in some scenarios of dynamical LMXB formation in GCs. The old, metal-rich, red population of GCs in NGC 3115 on the 
other hand is {\it at least} three times as efficient at creating LMXBs as the old, metal-poor, blue clusters. These data suggest that the higher formation efficiency of LMXBs in the red GC subsystems of many galaxies is largely a consequence of their higher metallicity. 
A comparison of the densities of field LMXBs in different galaxies does not reveal an obvious correlation with the age of the field stars
 as predicted by models in which the  LMXB formation rate in the field drops monotonically with time after an initial burst. This suggests that either a significant fraction of the field  LMXBs are created in
 GCs and subsequently injected into the field, or the LMXB formation rate has a more complex  time evolution pattern.  
 
\end{abstract}

\keywords{galaxies:general --- galaxies:individual(NGC 4365, NGC 3115) --- galaxies:star clusters --- globular clusters:general --- X-rays:binaries --- X-rays:galaxies}

\section{Introduction}

The hard X-ray emission from elliptical and S0 galaxies has long been  suspected of being associated with LMXBs (Trinchieri \& Fabbiano 1985).
With the advent of the {\it Chandra} X-ray Observatory, it is
 now possible to directly identify and study LMXBs in external  galaxies such as NGC 4697 (Sarazin, Irwin \& Bregman 2000) and NGC 4472 (Kundu, Maccarone \& Zepf 2002, hereafter KMZ; Maccarone, Kundu \& Zepf 2003, hereafter MKZ).

 LMXBs are formed especially efficiently in GCs, most likely due to dynamical interactions in the core (Clark 1975; Fabian et al. 1975). In the Galaxy, GCs account for $\lesssim$0.1\% of the stellar mass, but harbor $\sim$10\% of the L$_X$$\gtrsim$10$^{36}$ erg s$^{-1}$ LMXBs (e.g. Verbunt 2002). {\it Chandra} studies reveal that an even higher fraction of LMXBs in early type galaxies reside in GCs, ranging from 70\% in NGC 1399 (Angelini, Loewenstein \& Mushotzky 2001), to 40\% in NGC 4472 (KMZ, MKZ), and a lower limit of 20\% in NGC 4697 (Sarazin et al. 2000).  

	One of the surprising conclusions from our study of the GC-LMXB connection in NGC 4472 (KMZ) is that red GCs are $\approx$3 times
 more likely to host LMXBs than blue ones. Although a similar correlation is known to exist for the Galaxy (e.g. Grindlay 1987), the small number of LMXBs and clusters in the Galaxy makes it difficult to gauge the relative importance of various correlated GC properties. The rich GC systems of early type galaxies and the wide range of GC properties in different galaxies  allow us to isolate and understand the physical characteristics of GCs that drive LMXB formation. For example, the red colors of GC systems
 identified in broadband optical filters largely reflect a metal-rich population, but in different galaxies these clusters can also be younger than the blue GCs by varying amounts (e.g. Puzia et al. 2002; hereafter P02).  It has also been argued that LMXBs may preferentially be created in younger GCs where the larger mass of the turnover stars increases the interaction rate in GCs (e.g. Di Stefano et al 2002. Also see Rappaport et al. 2001). The discovery of a significant population of intermediate age GCs in NGC 4365 by P02 provides a unique sample to test for any effect of age on the LMXB formation rate. Furthermore, the P02 infrared study yields the only data set that constrains the ages and metallicities of large numbers of GCs in the  inner regions of galaxies where the LMXB density is the  highest. 

In this Letter we compare the X-ray properties of  GCs in NGC 3115 and NGC 4365 with the age and metallicity constraints placed by the P02 observations  in order to isolate the primary physical property that drives the efficient
 formation of LMXBs in red GCs.   We also  probe the field LMXBs in early type galaxies to ascertain whether they probe the age of the field stars.

\section{Observations \& Data Analysis}

We  have analyzed the 36 ks archival {\it Chandra} ACIS-S3 image of NGC 3115 observed on 2001, June 14 (PI: Irwin) and the 41 ks image of NGC 4365 observed on 2001 June 2 (PI: Sarazin). 
 Using the procedure laid out in KMZ and MKZ we identify 149 X-ray point sources in NGC 4365 and 90 objects in NGC 3115 in the 
0.5-8 keV band. The bulk of these sources are LMXBs. Only 10-15 of the candidates in each galaxy are expected to be contaminating objects such as background AGNs (Brandt et al. 2000, Mushotzky et al. 2000).  Neither of these  galaxies shows evidence of significant emission from bright, hot gas that might affect point source detection even in the innermost regions of the galaxies.  

The {\it HST}-WFPC2 V and I band observations of NGC 3115 and NGC 4365 have been analyzed by us in Kundu \& Whitmore (1998; 2001a) and P02. P02 have also analyzed ground-based K-band images of these galaxies. We refer the reader to these papers for details of the data reduction and analysis of their GC systems.  We achieved   0.3$''$ r.m.s. relative astrometric accuracy between the optical and X-ray sources by bootstrapping the {\it HST} and {\it Chandra} positions 
using the procedure outlined in MKZ. We adopt a distance modulus of 29.93 mag for NGC 3115 and 31.55 mag for NGC 4365 (Tonry et al. 2001).

\section{The LMXBs in Globular Clusters }

	In order to determine the fraction of LMXBs in GCs we consider only the X-ray sources that lie within the WFPC2 field of view.  36 of the LMXB candidates in NGC 3115 and 44 in NGC 4365 meet this criterion (we eliminated the source corresponding to the nucleus of each galaxy which may be associated with their central black hole). Nine of these LMXBs in NGC 3115 and 23 in NGC 4365 are within 0.5$''$ of a GC and are considered to be matches. There is a natural break in the LMXB-GC angular separation at this distance and no further candidates are added in either galaxy when the matching radius is increased to 0.7$''$. Thus 25\% of the LMXBs in NGC 3115 are associated with GCs, and 40\%  in NGC 4365. 

	The optical color distributions of GCs in many early type galaxies are known to be bimodal (e.g. Kundu \& Whitmore 2001a, 2001b). This is usually attributed to differences in the metallicity distributions of the subpopulations of old GCs. While NGC 3115 has a bimodal color distribution (Kundu \& Whitmore 1998), NGC 4365 is one of the few galaxies with a confirmed unimodal distribution (Kundu \& Whitmore 2001a).
In each bimodal GC system where the LMXB population has been studied -e.g. the Galaxy (Bellazzini et al. 1995) and NGC 4472 (KMZ)- the red GCs have a significantly higher fraction of LMXBs than the blue ones. 

	The color-magnitude diagrams of the NGC 3115 and NGC 4365 GCs are plotted in Fig 1, with LMXB hosts represented by filled symbols. It is apparent that the red GCs in NGC 3115 are preferred sites for LMXB formation. 
We showed in Kundu \& Whitmore (1998) that there are roughly equal numbers of
red and blue GCs in the WFPC2 image of NGC 3115 with a dividing color between 
V-I=1.03 and  V-I=1.06. Any choice of partition between red and blue GCs 
within this color range places 8 of the 9 GC-LMXBs in the red GC system of NGC 3115. Even if one were to conservatively assign the GC-LMXB with V-I=1.07 that is near the cusp of the color distribution to the blue population,
the efficiency of LMXB formation in red GCs is at least 3 times that in blue GCs.  On the other hand  the GC-LMXBs in NGC 4365 show no strong preference for redder GCs. The $<$V-I$>$=1.10$\pm$0.03 mean color 
of the GC-LMXBs  is only marginally redder than the  $<$V-I$>$=1.06$\pm$0.01 value of the 546 GCs within the color range 0.6$<$V-I$<$1.4. 

\section{Discussion}
\subsection{Are LMXBs Formed more Efficiently in Young Globular Clusters?}

 	Using K-band observations  to break the age-metallicity degeneracy that plagues optical colors, P02 showed that the broad optical color distribution of NGC 4365 GCs is due to the presence of a significant fraction ($\approx$50\% in the field of view) of intermediate age ($\sim$5 Gyr) GCs. This interpretation has been confirmed by spectroscopic observations of a handful of GCs by Larsen et al. (2003). P02 also 
showed that both the metal-poor and the metal-rich GCs in NGC 3115 appear to be old ($\sim$12 Gyr). In Fig 2 we plot the V-K vs V-I distribution of the GCs (and GC-LMXBs) in NGC 4365 and NGC 3115. 
While the age of an individual GC is only constrained to several Gyrs and [Fe/H] to a few tenths of a dex,  the differences in the overall GC system properties of the two galaxies, in particular the intermediate age population of NGC 4365, can clearly be seen. We refer the reader to P02 for details. Note that the P02 observations are largely limited by the depth of the K-band data. Thus, only a subsample of the Fig 1 GCs 
have age-metallicity constraints. 

	Figure 2 reveals that all the GCs in the P02 sample that host LMXBs in NGC 3115 are metal-rich and have solar or higher metallicity. Given the old ages ($\gtrsim$10 Gyr)
of both the red and blue GCs in this galaxy our observations strongly suggest that metallicity is a significant parameter that drives the formation of LMXBs in GCs. It appears likely that the higher efficiency of LMXB formation in the red GCs of galaxies with clearly bimodal optical color distributions such as NGC 4472 (KMZ) is also driven by metallicity effects. On the other hand, the GC system of NGC 4365 which spans a large range of ages does not reveal any obvious correlation of LMXB efficiency with GC age. Specifically, there is no evidence that intermediate age GCs are more likely to host
LMXBs. Although the specific choice of age separating the intermediate age and old GCs is fairly arbitrary, a typical choice of 5 Gyr actually gives a slightly {\it higher} fraction of LMXBs in old GCs ($\approx$20\%) than in young ones ($\approx$10\%). 

However, there are strong selection effects due to the K-band photometric limits (P02), coupled with the fact that younger GCs are brighter, and the strong 
preference of LMXBs to form in the brightest GCs (Fig 1 \& KMZ). We showed in KMZ that the luminosity effect implies that roughly equal number of LMXBs are formed
per unit GC mass at all luminosities. Thus, in order to fairly judge the
effect of age and/or metallicity one should compare the LMXB formation efficiency for GCs of equal mass. To this end we used the ages and metallicities of individual GCs and the implied Bruzual \& Charlot (2000) model luminosities to estimate the masses of the GCs. We note here in passing that the
LMXB fraction per unit mass of GCs in our sample is 10$^{-7}$ M$_{\hbox{$\odot$}}$$^{-1}$ and indeed roughly constant for all GC masses.

	Figure 3 shows the [Fe/H] vs. mass distribution of GCs and GC-LMXBs in NGC 3115. In conjunction with Figs 1 and 2 it is apparent that the red GCs are favored sites for LMXB formation due to the higher metallicities of these clusters. Figure 4 plots the age and metallicity vs. mass  distributions of the NGC 4365 GCs. The lower panel of Fig 4  shows no strong correlation of LMXB formation rate with GC metallicity in NGC 4365. It is not clear if this is due to selection effects, such as the fact that the K-band data undersample the fainter metal-poor GCs in this more 
distant galaxy, or possible physical processes e.g. the metallicity dependence is not continuous but a threshold effect above which the efficiency of LMXB formation increases. More significantly, the upper panel of Fig 4 shows no evidence that younger GCs form LMXBs more efficiently. We attempted to separate the sample into young and old populations using a range of dividing ages from 4-8 Gyr (see P02) to search for any evidence of a higher LMXB formation rate in young GCs.
For a dividing age of 8 Gyr the LMXB density is 1.1($\pm$0.4)$\times$10$^{-7}$ M$_{\hbox{$\odot$}}$$^{-1}$ for the young GCs and  0.7($\pm$0.5)$\times$10$^{-7}$ M$_{\hbox{$\odot$}}$$^{-1}$  for the old ones. Conversely, dividing the populations at 4 Gyr yields 0.7($\pm$0.5)$\times$10$^{-7}$ M$_{\hbox{$\odot$}}$$^{-1}$ for the young population and  1($\pm$0.4)$\times$10$^{-7}$ M$_{\hbox{$\odot$}}$$^{-1}$ for the old GCs.  Furthermore, we point out that if the efficient formation of LMXBs in the red GCs of
 the Galaxy and M31 is due to their $\sim$2 Gyr younger age compared to 
the blue clusters, we would expect to see a significantly larger
 difference in LMXB formation efficiencies between the young and the old GCs in NGC 4365 due to their larger age spread. Given that roughly half the GCs in our WFPC2 image are young (P02) our data clearly does not support this premise. Also, the deep V-I observations indicate that $\approx$3\% of the NGC 4365 GCs host LMXBs  which is similar to the $\approx$4\%  fraction observed in other galaxies (KMZ). 

	Thus we conclude that the efficiency of LMXB formation in a GC shows no obvious correlation with the age of the GC. However, the larger fraction of LMXBs seen in the red GC systems of NGC 3115, and likely the other galaxies with clearly bimodal GC color distributions,
appears to be related to the higher metallicities of these clusters. There is no convincing theoretical model that explains this metallicity effect. Systematic variations as a function of metallicity in the dynamical properties of clusters, the dynamical binary formation rates, or the efficiency of mass transfer in LMXBs are all possibilities at this time. These observations lay the
preliminary observational goalposts for understanding the formation of LMXBs in GCs. Deeper optical and K-band observations of large samples of extragalactic GCs will clearly be invaluable in probing the observational signatures in more detail. 

\subsection{Do Field LMXBs Trace the Age of Field Stars?}


	Models of LMXB formation due to the evolution of binaries in the field  suggest that the rate of formation of bright LMXBs decreases monotonically after an initial peak (White \& Ghosh 1998; Wu 2001). If this is the case, the properties of the field LMXBs may be used
probe the star formation history of the host galaxy. In this scenario the vast majority of the bright LMXBs in the Milky Way must be associated with recent star formation in the Galactic disk. However, we have discovered 27 field LMXBs with L$_X$ $>$ 3$\times$10$^{36}$ ergs/s within the WFPC2 image of NGC 3115 and, we estimate, $\approx$50 in the entire ACIS-S3 frame. There are roughly twice as many LMXBs of comparable luminosity in the Galaxy.
 Thus, one would either require continuous star formation in 
a region of NGC 3115 with mass comparable to the Milky Way disk, or a significant star formation 
episode in the recent past to account for the field LMXBs in NGC 3115. There is no clear evidence for recent star formation in either the GCs or the field stars
  in NGC 3115 (P02), strongly suggesting that the LMXB formation rate does not correlate trivially with the age of the stellar system.

To further search for age effects we plot the local specific frequencies, S$_N$, of GCs, GC-LMXBs and field LMXBs of NGC 3115, NGC 4365, NGC 1399
\& NGC 4472 as a function of host galaxy absolute magnitude in Fig 5. The local S$_N$ is defined as N$\times$10$^{0.4(M_{V(FOV)}+15)}$, where N is the number
of relevant candidates and $M_{V(FOV)}$ is the absolute magnitude of the host galaxy within the WFPC2 field of view.  LMXBs fainter than 2$\times$10$^{37}$ ergs/s have been eliminated for the nearby NGC 3115 system in order to minimize selection bias. 
The LMXB values have been arbitrarily multiplied by 20 to facilitate comparison with the GC S$_N$s.

	Figure 5 reveals that broadly the GC and GC-LMXB specific frequencies track each other, which essentially reiterates our suggestion (KMZ) that the overall efficiency of LMXB formation in GCs is similar in different galaxies. The S$_N$ of field LMXBs also appears to be similar in the inner regions of these galaxies.
However, if all LMXBs brighter than $\gtrsim$2$\times$10$^{36}$ ergs/s are formed in regions with continuous or recent bursts of star formation as suggested by Wu (2001), NGC 4365 which is likely to have a component of intermediate age stars in the field  should have a higher value of S$_{N(Field)}$. The approximately several gigayear delay in the peak of the LMXB formation rate  suggested by White \& Ghosh (1998) should have 
further enhanced the field LMXB density in NGC 4365 at the present epoch. No obvious enhancement is seen in Fig 5.

Thus, we conclude that our present data set does not support a scenario in which the LMXB formation efficiency decreases monotonically after an initial peak. It is possible that a large fraction of present day LMXBs reflect an old stellar population where Roche lobe overflow in a moderately tight binary is triggered by the evolution of donor star off the main sequence (e.g. Podsiadlowski, Rappaport \& Pfahl 2002). Alternately a significant fraction of field LMXBs may actually have been created in GCs and subsequently injected into the field (MKZ) either by GC destruction (e.g. Vesperini 2000; 2001) or dynamical kicks (e.g. Phinney \& Sigurdsson 1991). 
  Finally we note, that in much of this Letter we have implicitly assumed that our snapshot study accurately reflects 
the relative densities of various populations of LMXBs. Followup {\it Chandra} observations are necessary to quantify possible differences in the transient LMXB populations (Piro \& Bildsten 2002) that may affect these conclusions.  
  
	This research was supported by NASA LTSA grants
NAG5-11319 and NAG5-12975, and Chandra grant AR3-4010X. THP acknowledges support from \emph{Deut\-sche For\-schungs\-ge\-mein\-schaft, DFG\/} project \# Be~1091/10--2.

\clearpage

\begin{figure*}[!ht]
\centerline{\psfig{figure=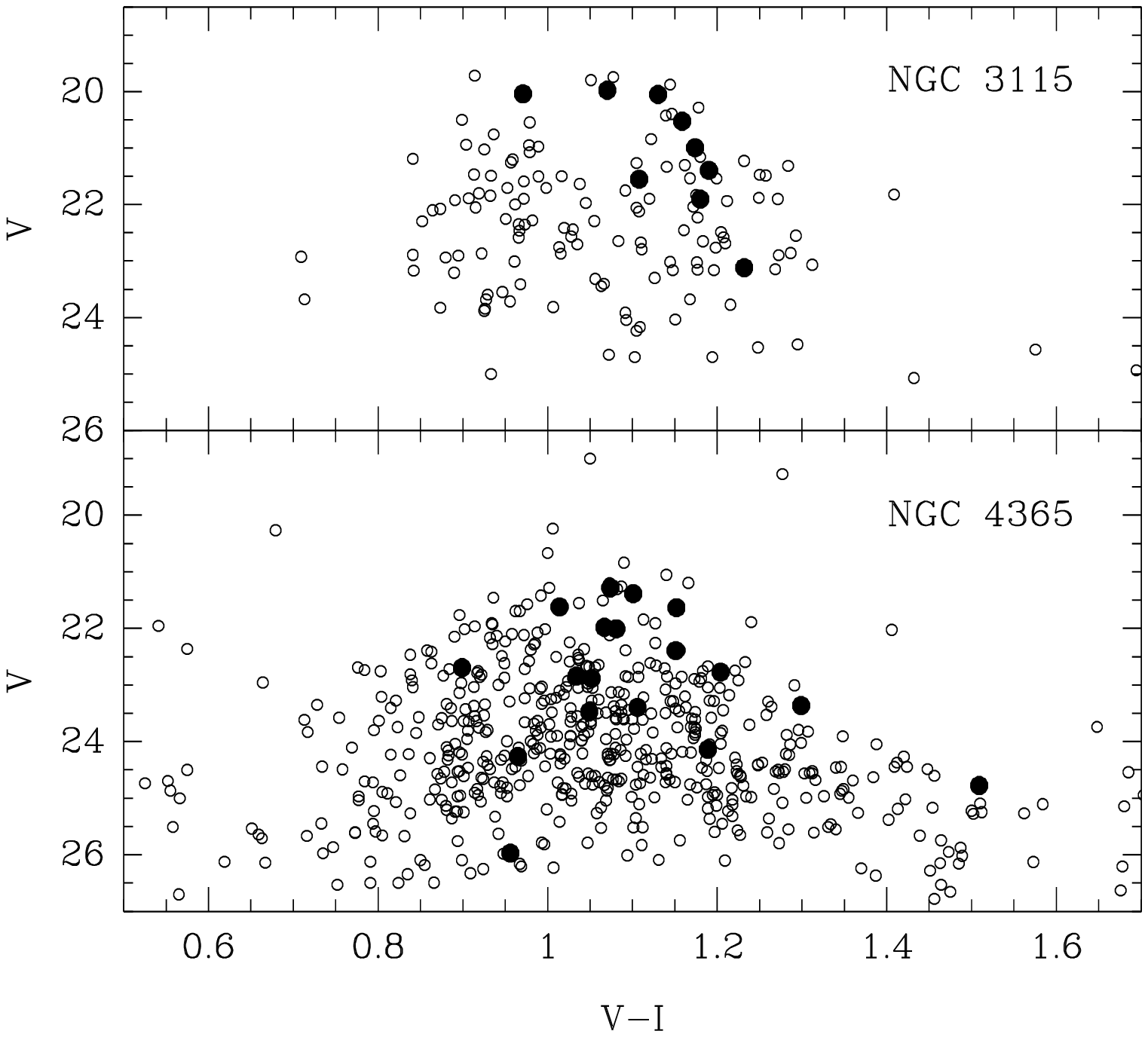,width=10cm,angle=0}}
\caption{ Color-magnitude diagrams for the globular cluster candidates in NGC 3115 and NGC 4365. Filled points represent clusters with LMXB counterparts.  }
\end{figure*}

\begin{figure}
\centerline{\psfig{figure=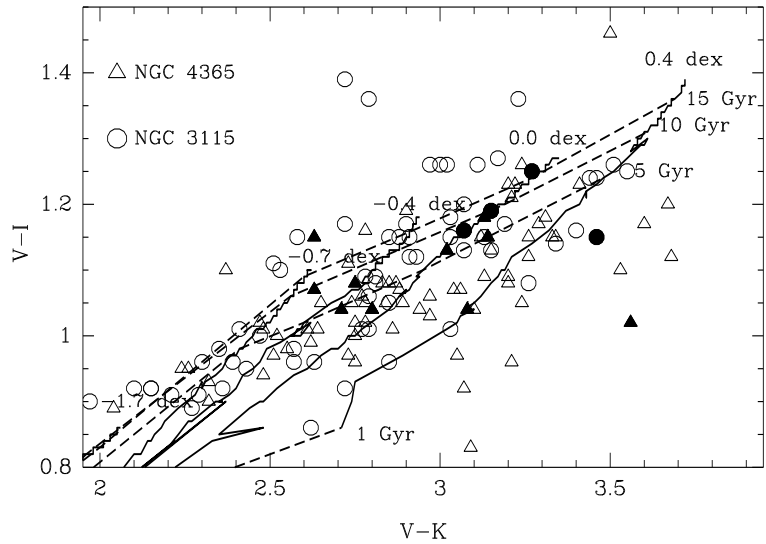,width=10cm,angle=0}}
\caption{V-I vs V-K color-color plot of the globular clusters in NGC 4365 and NGC 3115 from P02. Dashed and solid lines indicate Bruzual \& Charlot (2000) isochrones and isometallicities for various fiducial values, assuming a Salpeter IMF. Filled points mark GC-LMXBs. }
\end{figure}

\begin{figure}
\centerline{\psfig{figure=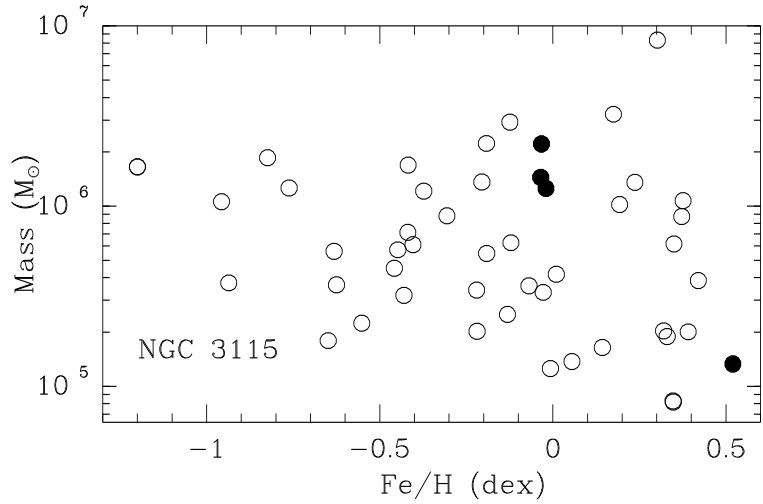,width=10cm,angle=0}}
\caption{Metallicity vs. mass distribution for the globular clusters in NGC 3115 using the optical and IR colors of P02 and the Bruzual \& Charlot (2000) models. Filled points represent GC-LMXBs. Typical uncertainties in [Fe/H] of individual GCs are several tenths of dex (See P02).  }
\end{figure}

\begin{figure}
\centerline{\psfig{figure=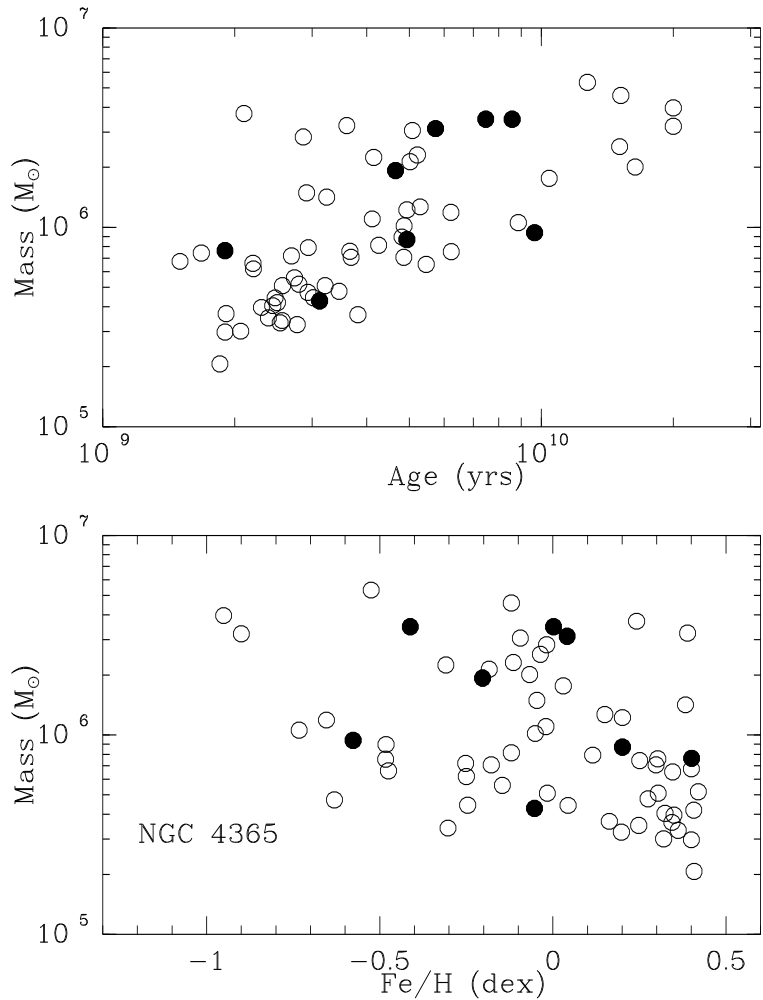,width=10cm,angle=0}}
\caption{Top: Age vs. mass  for the NGC 4365 globular clusters. The uncertainty in the age of an individual GC is a few Gyrs (See P02). Bottom: Metallicity vs. mass distribution. }
\end{figure}

\begin{figure}
\centerline{\psfig{figure=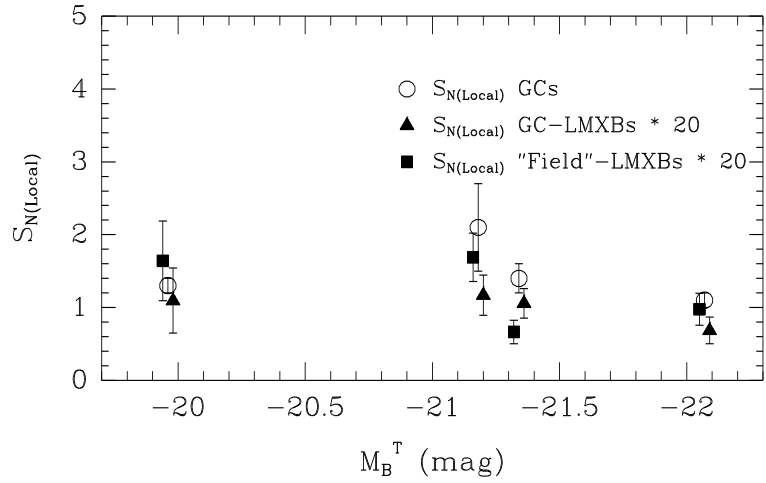,width=10cm,angle=0}}
\caption{ The local specific frequency of GCs, GC-LMXBs, and field LMXBs in (from left to right) NGC 3115, NGC 4365, NGC 1399 and NGC 4472, as a function of the total B-band absolute magnitude of the galaxies. Arbitrary small offsets along the X-axis have been applied to the S$_N$s for clarity.}
\end{figure}

\end{document}